\documentclass{llncs}
\usepackage{amssymb}
\usepackage{graphicx}
\usepackage{csvsimple}
\usepackage{caption}
\usepackage[figure]{algorithm2e}
\graphicspath{ {images/} }

\begin{document}
\mainmatter              
\title{Algorithms and Architecture for Real-time Recommendations at
  News UK}
\titlerunning{Real-time News Recommendations}  
%
\author{Dion Bailey\inst{1} \and Tom Pajak\inst{1} \and
  Daoud Clarke\inst{2} \and Carlos~Rodriguez \inst{3}}
\authorrunning{Dion Bailey et al.} 
%
\tocauthor{Dion Bailey, Tom Pajak, Daoud Clarke and Carlos Rodriguez}
\institute{News UK, London, UK,\\
\email{\{dion.bailey, tom.pajak\}@news.co.uk},\\
\texttt{https://www.news.co.uk}
\and
Hyperparameter Ltd., London, UK,\\
\email{daoud@hyperparameter.com},\\
\texttt{http://www.hyperparameter.com/}
\and
Kainano Ltd., London, UK\\
}

\maketitle              

\begin{abstract}
Recommendation systems are recognised as being hugely important in
industry, and the area is now well understood. At News UK, there is a
requirement to be able to quickly generate recommendations for users
on news items as they are published. However, little has been
published about systems that can generate recommendations in response
to changes in recommendable items and user behaviour in a very short
space of time. In this paper we describe a new algorithm for updating
collaborative filtering models incrementally, and demonstrate its
effectiveness on clickstream data from \emph{The Times}. We also describe
the architecture that allows recommendations to be generated on the
fly, and how we have made each component scalable. The system is
currently being used in production at News UK.
\end{abstract}
\section{Introduction}

Two years ago News UK completed a refresh of our data platforms and
brought this process in-house. Now that we had greater access to our
data we wanted to provide our customers with a premium digital
experience based on their individual habits and behaviours. Other
competitors in the market offer this, and we believe it will help us
improve engagement and reduce churn. There are a number of products in
the market that attempt to achieve this, however, we required a
platform that was able to adapt to the constant changing news cycle
and not centred around evergreen or e-commerce data. The decision was
made to develop a platform tailored to our unique business models. We
have two major News Titles with two different business models,
\emph{The Times \& The Sunday Times} and \emph{The Sun}.  On the
Times, we are in the unique position of knowing a lot about our users,
their behaviours, their preferences and their level of engagement with
our products due to the digital product suite being behind a
paywall. \emph{The Times \& The Sunday Times} currently has over
400,000 subscribers. \emph{The Sun} is a brand that is going for major
reach across all of its products; since removing its paywall last year
it has become the second largest UK newspaper \cite{ponsford2017}.

The platform we designed is intended to improve the numbers above and
increase retention by employing personalization techniques where it
makes sense for users. We want to maintain the current editorial
package, but use the information we have at our disposal to present
content relevant to the user, which provides a more engaging product
experience.


Our paper is structured as follows. In the remainder of this section
we outline related work, our approach and the datasets used. In Section
\ref{section:algorithms} we describe the algorithms we used and the
offline evaluations we performed. This includes the major contribution
of this paper, a novel algorithm for updating collaborative filtering
models incrementally. In Section \ref{section:evaluation} we describe
our evaluation, and in Section \ref{section:results} we give our
results. We show that the incremental approach works as well as
non-incremental under the right conditions. In Section
\ref{section:architecture} we describe the architecture of our system,
concluding in Section \ref{section:conclusion}.


\subsection{Related Work}

Diaz-Aviles et al.~\cite{diaz2012real} describe an algorithm for
real-time recommendations called Stream-Ranking Matrix Factorization
(RMFX) in the context of recommending for social media. This performs
matrix factorization and ranking of recommendations on streaming
data. However their system requires specifying the set of users and
items in advance, which is not appropriate in our setting where we
must handle new users and items (in our case new articles) all the
time.

The xStreams system of Siddiqui et al.~\cite{siddiqui2014xstreams}
does handle new users and items, however it does not incorporate the
matrix factorization algorithms which provide the current state-of-the
art recommendations.

The system used for real-time recommendations on YouTube is described
by Covington et el.~\cite{covington2016deep}, which uses a
sophisticated deep-learning model built on TensorFlow
\cite{tensorflow2015whitepaper}. However the focus of the paper is on
the deep learning model used rather than the real-time aspect of the
system.

\subsection{Approach}

Our approach has been a pragmatic one. We performed an offline
evaluation of a number of standard approaches to recommendation
generation. We then chose the best performing systems to implement in
production. The initial requirement for recommendations was to send an
email to users once a day with personalised recommendations. Our
first implementation precomputed recommendations for all
users. These were then sent to users via email at a preconfigured
time. We found that it took a long time to precompute and store the
recommendations for all users: with a Spark cluster of forty machines,
the recommendations could take up to half an hour to generate.

We decided to re-architect our system when we were tasked with
building a system to serve recommendations on demand for \emph{The
  Times} and \emph{The Sun} websites. In particular, for \emph{The
  Sun}, new content is generated throughout the day, and we wanted that
content to be recommendable as soon as possible. In our new
architecture, we no longer precompute recommendations. Instead, models
are updated continuously as new information about users is
received. The models are stored in a database that allows them to be
very quickly retrieved, and recommendations are generated at the point
that they are needed via an HTTP request to our API.

This approach not only means that recommendations are always
up-to-date, but also means that we do not need to generate and store
recommendations for all users, a process which is both time and space
intensive.

\subsection{Datasets}

We collected ten days' worth of \emph{Times} user behaviour from web
logs. For each pair of user and item, we collated the user events as
follows:
\begin{itemize}
  \item \textbf{Dwell time:} this was estimated based on time between
    subsequent clicks. If the time between subsequent clicks by the
    same user was less than 30 minutes, it was assumed that the user
    spent that time on the first item clicked on. Thus the last item
    clicked on by a user would never receive a dwell time event.
  \item \textbf{Shares:} the number of times that the user shared the
    item.
  \item \textbf{Comments:} the number of times the user commented on
    the item.
\end{itemize}
These data were translated using a simple rule to determine whether or
not there was an implicit expression of interest in the item by the
user. We call such interactions ``significant''. We defined a
significant interaction to be a dwell time of more than ten seconds,
or any positive number of shares or comments. Reducing the data to
this simple binary signal simplified the choices we had to make in
designing and evaluating the algorithms. We plan to investigate more
sophisticated possibilities in future work.

\section{Algorithms}
\label{section:algorithms}

We evaluated two recommendation algorithms, one collaborative in
nature and one content based, and two baselines that we wished to
improve upon: global popularity ranking and randomly chosen
articles. The first chooses the articles that have the highest number
of significant actions in the training set, and the second chooses
from articles seen in the training set at random.

\subsection{Incremental Updates for Collaborative Recommendations}

\begin{algorithm}
  \SetKwData{Users}{users}
  \SetKwData{Items}{items}
  \SetKwData{Ratings}{ratings}
  \SetKwData{R}{R}
  \SetKwData{X}{X}
  \SetKwData{Y}{Y}
  \SetKwFunction{LatentFactorUpdate}{LatentFactorUpdate}
  \Users $\leftarrow$ empty dictionary\;
  \Items $\leftarrow$ empty dictionary\;
  \Ratings $\leftarrow$ empty dictionary\;
  \While{more batches exist}{
    read new batch with $n_u$ users and $n_i$ items\;
    \ForEach{user $u$ in batch}{
      \tcp{initialise unseen user vectors to a new random vector:}
      \If{$u$ not in \Users}{
        $\Users[u] \leftarrow $ new initial user vector\;
      }
      \tcp{keep track of all user ratings so far:}
      \eIf{$u$ not in \Ratings}{
        $\Ratings[u] \leftarrow $ empty set\;
      }{
        $\Ratings[u] \leftarrow \Ratings[u] \cup \textrm{new items
          with significant actions for $u$ in this batch}$\;
      }
    }
    \tcp{initialise unseen item vectors to a new random vector:}
    \ForEach{item $i$ in batch}{
      \If{$i$ not in \Items}{
        $\Items[i] \leftarrow $ new initial item vector\;
      }
    }
    \tcp{perform a learning iteration:}
    $\R \leftarrow \textrm{matrix of shape $n_u \times n_i$ with values from \Ratings}$\;
    $\X \leftarrow \textrm{matrix of shape $n_u \times k$ with values from \Users}$\;
    $\Y \leftarrow \textrm{matrix of shape $n_i \times k$ with values from \Items}$\;
    \LatentFactorUpdate{\R, \X, \Y}\;
    update \Ratings, \Users and \Items with values from \R, \X and \Y\;
  }
  \vspace{0.3cm}
  \caption{Incrementally updating a collaborative filtering
    model. Each batch is a streamed collection of user actions. The
    positive integer $k$ is a parameter of the underlying
    collaborative filtering algorithm specifying the dimensionality of
    the factorization. The matrices \textsf{X}, \textsf{Y} and
    \textsf{R} are built from dictionaries of ratings, users and items
    by defining an order on users and items and iterating the maps in
    that order. The function \texttt{LatentFactorUpdate} updates
    \textsf{X} and \textsf{Y} from \textsf{R} using the underlying
    collaborative filtering algorithm.}\label{algorithm}
\end{algorithm}

Our main contribution is an algorithm for updating collaborative
models incrementally, described in Figure \ref{algorithm}. In theory,
the model could work with any collaborative filtering algorithm that
allows user and item vectors to be updated from some initial state.

\begin{figure}[t]
  \begin{center}
    \includegraphics[scale=0.5]{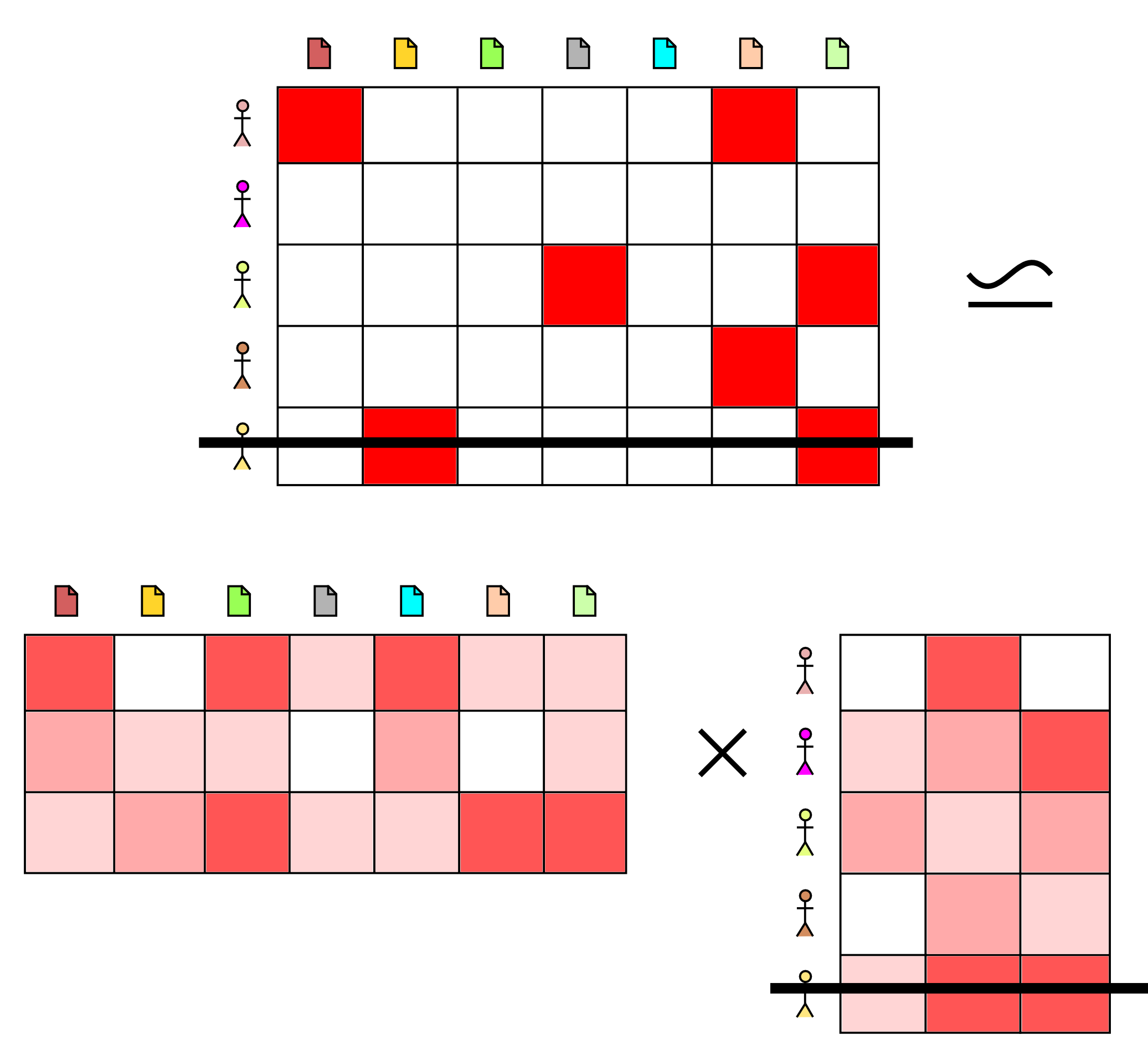}
  \end{center}
  \caption{Depiction of the matrix factorization algorithm, and how a
    user can be added or removed by adjusting the matrices. The matrix
    at the top depicts the matrix of user preferences, with users in
    rows and items in columns. This is approximately factored into two
    matrices, one for items and one for users. Removing a user from
    the preference matrix results in removing the corresponding row in
    the users matrix. In our algorithm we do the converse by adding a
    randomly initialised vector to the users matrix along with the
    user's actions to the preference matrix.}\label{decomposition}
\end{figure}

The algorithm is based on the observation that in collaborative
filtering algorithms that decompose a matrix into products of latent
factors, the set of users and items under consideration can easily be
altered by adding or removing rows or columns from the matrices (see
Figure \ref{decomposition}). It works by processing batches of user
actions, updating the vectors only for users in each batch. We found
that it was necessary to keep track of all user actions performed to
date; for each user in the batch, we retrieve these actions and
perform a model update using the underlying collaborative algorithm.

Some concerns when implementing this algorithm are:
\begin{itemize}
  \item determining the size of the batch: if batches are too small,
    the algorithm does not work reliably. This was mitigated by adding
    in some randomly chosen users if there were not enough in the
    batch. The number of users needed in the batch was determined
    empirically.
  \item processing batches in parallel: if the level of parallelism is
    too high, the algorithm does not converge to an optimal solution.
\end{itemize}

Storing and retrieving user actions and user and item vectors becomes
a major concern when implementing this algorithm at scale. Most of our
effort has been around designing an architecture to do this reliably,
described in Section \ref{section:architecture}.

\subsection{Algorithmic Complexity}

Keeping the model constantly up-to-date comes with an associated
computational cost. Let $f(n)$ be a function describing the time
complexity of the underlying matrix factorization algorithm, where $n$
is the number of user-article pairs in the dataset, i.e.~the number of
non-zero entries in the matrix. Our algorithm splits the $n$
datapoints into batches of size $b$. In the worst case (when there is a very
small number of users), all the data from the previous batches needs
to be included in each batch. Thus the complexity is
$$O(\sum_{i=1}^{n/b} f(ib))$$
In the case where $f$ is linear, this becomes $O(n/2 + n^2/2b) =
O(n^2/b)$. Making $b$ bigger mitigates the additional cost, and if $b
= n$ we recover $O(n)$ complexity.

\subsection{Collaborative System}

The collaborative filtering algorithm we use for the latent factor
update step is based on an approach to implicit feedback datasets
\cite{hu2008collaborative}. This suggests an alternative to treating
a rating matrix $R$ as a set of explicit ratings given by the
user. Each value is considered as a rating together with a confidence
in that rating. In their scheme, items with a low rating are given a
low confidence.

The implementation we use makes use of conjugate gradient descent to
perform the updates \cite{takacs2011applications}. This is a faster
algorithm with time complexity $O(nEk^2)$ where $E$ is the number of
iterations and $k$ the number of latent factors. The original
algorithm has cubic time complexity with respect to $k$ \cite{pilaszy2010fast}.

Intuitively, it makes sense to consider algorithms for implicit
feedback. This approach is in theory perfectly suited to our case, since
we have only implicit signals of interest in articles from users and
no reliable signal that the user is not interested in an
article. Thus, if we know that a user has shared an article or read it
for five minutes, we can be fairly confident that the user is
interested in it. However, if the user did not read the article we can
be much less certain that the user is not interested in it. In order
to verify this hypothesis, we performed some initial experiments in
which we found that algorithms designed for implicit signals gave much
higher accuracy on our datasets than those designed for explicit
signals.


We used Ben Frederickson's open source
implementation\footnote{\texttt{https://github.com/benfred/implicit}}
which is written in Python and Cython, which compiles to C. The
library makes use of the low-level BLAS library for vector operations,
which makes it very fast.

On top of this, we implemented the weighted regularization scheme
described in \cite{zhou2008large} in the Cython version of the
library, which we found provided a modest improvement in accuracy in
our preliminary experiments.

The algorithm assumes that for each user $u$ and item $i$ we have
values $p_{ui} \in \{0, 1\}$ which describes whether or not the user
has implicitly expressed a preference for the item and $c_{ui} \in
\mathbb{R}$ that expresses our confidence in the user's preference for
that item \cite{hu2008collaborative}. These values are assumed to be
derived from the implicit rating $r_{ui}$ given by the user for the
item in such a way that $p_{ui} = 1$ if and only if $r_{ui} > 0$, and
$c_{ui} = 1$ if and only if $r_{ui} = 0$, with $c_{ui} > 1$
otherwise. One suggestion is that $c_{ui} = 1 + \alpha r_{ui}$ for
some parameter $\alpha$ to be tuned. The intuition behind this is that
items for which we have no information from a user are treated as a
negative signal with low confidence, while implicit signals from a
user are treated as a positive signal with higher confidence. In
practice, the values are chosen to have this form to make the vectors
in the intermediate computations sparse.

Given these matrices, the goal is to learn vectors $x_u$ for each user
and $y_i$ for each item such that we minimize the following:
$$\min_{x_*, y_*} \sum_{u, i} c_{ui}(p_{ui} - x_u^Ty_i)^2 +
\lambda\left(\sum_u n_{x_u}\|x_u\|^2 + \sum_i
n_{y_i}\|y_i\|^2\right)$$ where $\lambda$ is a regularization constant
and $n_{x_u}$ and $n_{y_i}$ are the number of non-zero entries in the
vectors $x_u$ and $y_i$ respectively. This is solved using the
conjugate-gradient approach described in
\cite{takacs2011applications}.

In the experiments we report here we used a dimensionality of 50 for
the factor vectors; we found this worked well in initial experiments.

\subsection{Content-based Learning to Rank}

The content-based system is based on the Learning to Rank model
\cite{liu2009learning}, which treats the task of ranking pages as a
supervised learning problem. We consider two ``classes'' of articles:
those that interest the user and those that do not. Since we do not
make use of any explicit signals from users, for the purpose of
training a model, we identify these two classes with the following
sets of articles:
\begin{itemize}
\item The ``positive'' articles are those that the user has had a
  significant interaction with.
\item The ``negative'' articles are a random sample of articles that
  the user has not interacted with at all.
\end{itemize}
For each user, we train a logistic regression model on this data and
use it to rank articles as to their likelihood of being of interest to
the user.

\subsubsection{Implementation Details}

We use the Liblinear implementation \cite{fan2008liblinear} of the
logistic regression algorithm, and perform a search on the cost
parameter for each user, considering costs of 10, 20 and 50, having
found that costs below this range are rarely optimal. We perform cross
validation on the training set, choosing the value that gives the
highest F1 score.

\subsubsection{Features}

\begin{figure}[t]
  \begin{center}
    \includegraphics[scale=0.2]{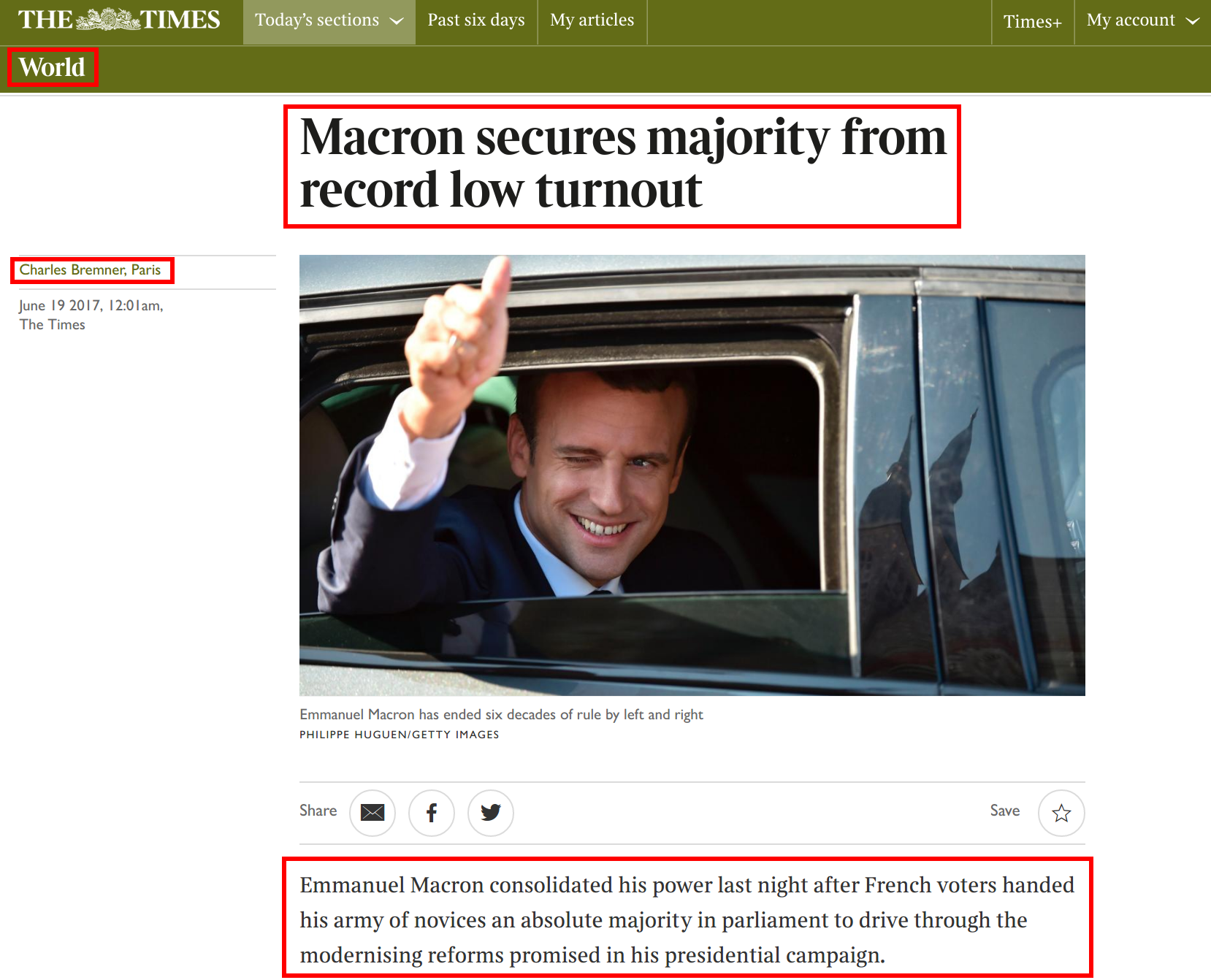}
  \end{center}
  \caption{A \emph{Times} article showing the features used for
    training the Content-based system (outlined).}
  \label{features-figure}
\end{figure}

We use the following features (see Figure \ref{features-figure}):
\begin{itemize}
\item The name of the article section,
\item Each term in the ``author'' text,
\item Each term in the article title and body.
\end{itemize}
All text is tokenized and lower-cased and a simple stop-word list is
applied, but no stemming or lemmatization is performed. Tokens from
each type of feature (section name, author text and title and body
text) are distinguished by adding a unique prefix for each type.
Features are treated as binary variables as is typical in document
classification. This means we look only at the presence or absence of a
feature, rather than counting occurrences.

\subsubsection{Feature selection}

We experimented with a variety of feature selection methods, but found
the following simple approach worked the best. We train a logistic
regression model using all features, then take the top 2,500 and
bottom 2,500 features (those with the highest and lowest coefficients)
from the learnt model, and set the remaining coefficients to
zero. This has the advantage of giving us a sparse model, which helps
to generate recommendations quickly since the model is smaller, but
also seems to provide a marginal improvement in recommendation
quality.

\subsubsection{Selecting Random Articles}

We found that when the ratio of irrelevant to relevant articles was
too high, the learning algorithm became unreliable. Restricting the
number of negative articles to at most five times the number of
positive ones provided a good compromise, representing the underlying
imbalance in the dataset whilst still keeping learning reliable.

\subsubsection{Boosting popular articles}

We found that the content-based system alone performed poorly compared
to the global top items baseline. Boosting the ranking of popular
items lead to a significant increase in our evaluation metric. To do
this, we computed a score
$$s_{ui} = p_{ui}(f_{i} + \beta)$$ where $p_{ui}$ is the probability
output by the logistic regression model for user $u$ and item $i$,
$f_{i}$ is the number of users that have a significant action for item
$i$ and $\beta$ is a smoothing value which we set at 10. The smoothing
value allows for items that have no significant actions to still be
recommended.

\section{Evaluation}
\label{section:evaluation}

\subsection{Evaluation metrics}

The offline evaluation method we have been using has been designed
with the following in mind:
\begin{itemize}
\item Make the evaluation representative of how the system will be used in
practice
\item Design the evaluation so that all types of algorithm can be compared
\item Make the evaluation metric intuitively simple
\end{itemize}
These considerations rule out the normal metrics that are used for
example in evaluating matrix factorisation algorithms, since we also
want to be able to evaluate classification based approaches. In our
first application, we send out an email containing 10 recommendations
to users. For this reason we have opted for the ``precision at 10''
metric.

\subsection{Evaluation procedure}

The data consists of (user, article, interactions) tuples, where
``interactions'' consists of all interactions that a user has had with
that article, including dwell time, comments and shares. We split these
data randomly into a training set consisting of 80\% of the data, and
the remaining 20\% is held out as a test set.  We then train a model
on the training set, generating recommendations for each user from
articles seen in the training set. For each user we take the top ten
recommendations excluding items that occur in the training set for
that user, and count the proportion of them that occur for that user
in the test set. This is averaged over all users to get the mean
precision at 10 score.

\section{Results}
\label{section:results}

\begin{figure}[t]
  \centering
  \begin{minipage}[t]{.49\textwidth}
    \centering
    \includegraphics[width=\linewidth]{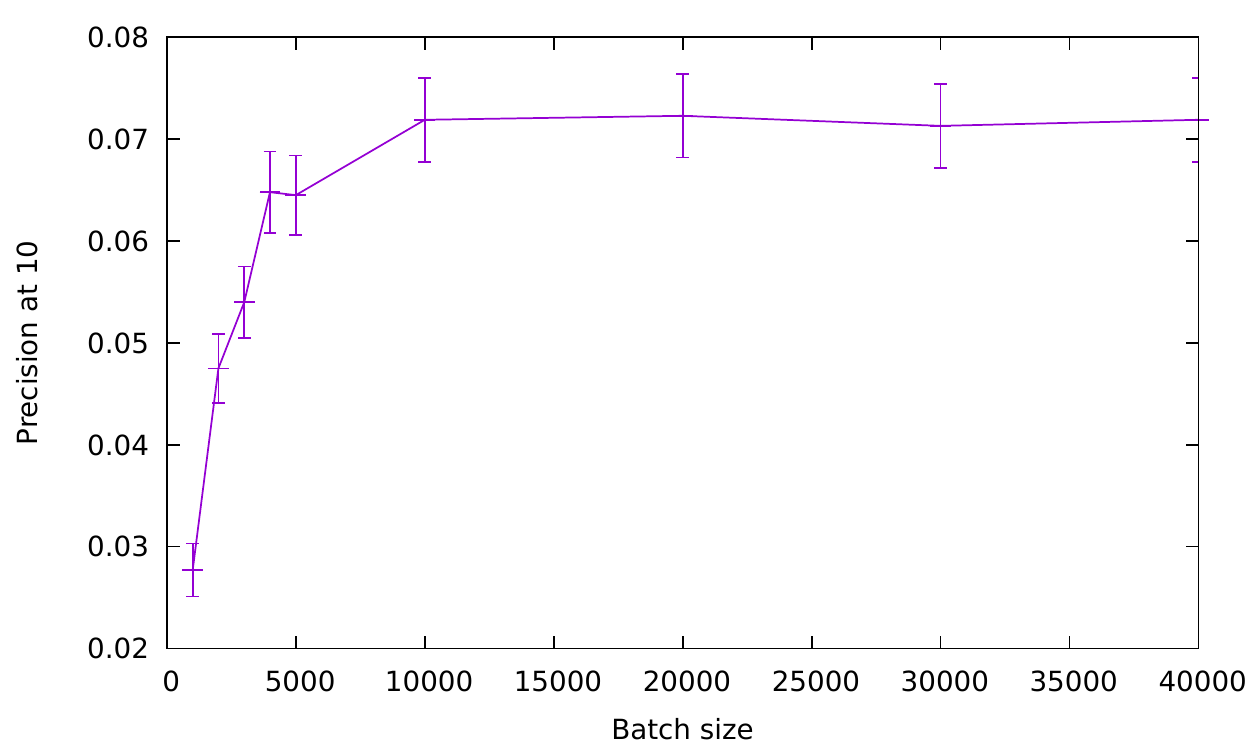}
    \captionof{figure}{Precision at 10 with batch size.}\label{results-batch-sizes}
    \vfill
  \end{minipage}\hfill%
  \begin{minipage}[t]{.49\textwidth}
    \centering
    \includegraphics[width=\linewidth]{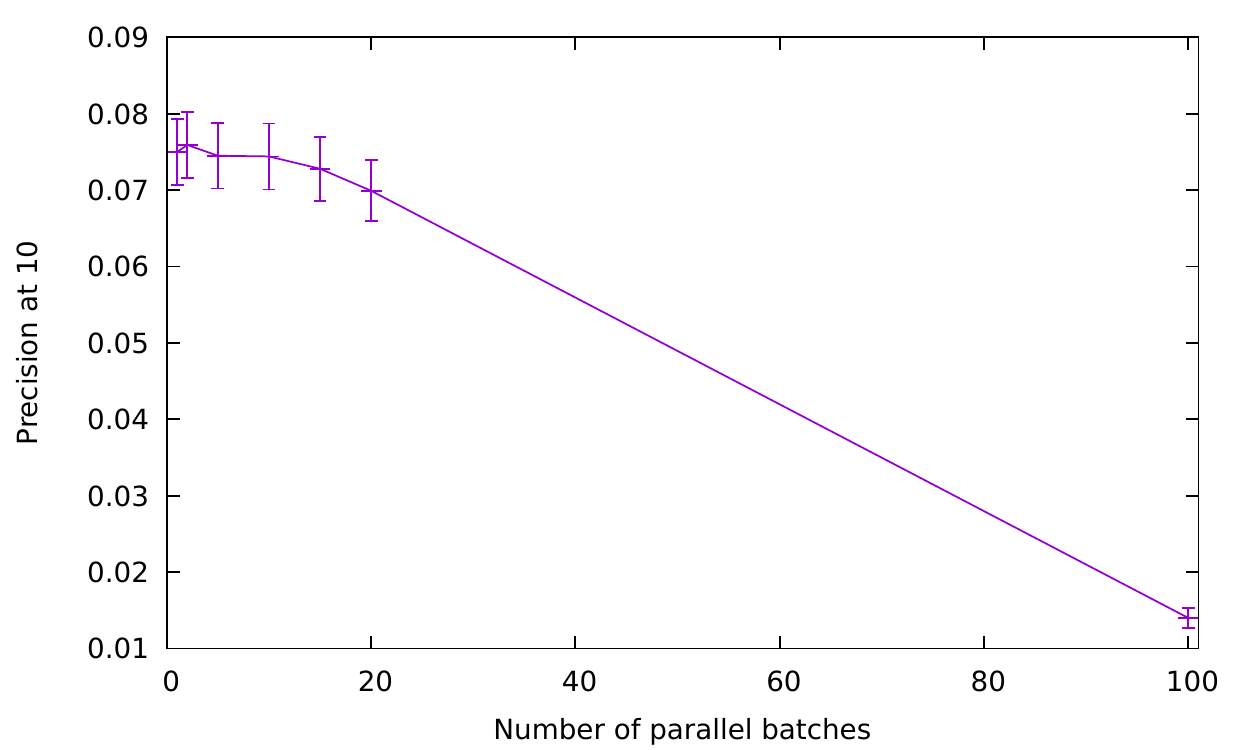}
    \captionof{figure}{How precision at 10 is affected by the number of batches that are
      processed in parallel, for a batch size of 100,000.}\label{results-parallel-batches}
  \end{minipage}
\end{figure}

Results for the systems we evaluated are shown in Table
\ref{results-systems}. Only the collaborative system was significantly
better than the Global Top Items baseline. Figure
\ref{results-batch-sizes} shows how the precision at 10 score varies
with batch size. For batch sizes over 10,000 there is no significant
increase in the score. Using a single batch is equivalent to training
using the underlying collaborative filtering model; since the score is
not significantly different to training with batches of 10,000, it is
clear that our incremental approach works, at least with the set-up we
have chosen. In initial experiments, we also found that the batch size
needed is dependent on the dimensionality of the learnt factors: the
higher the dimensionality, the larger the batch size needed to avoid
harming the score. One limitation of our approach is that the batch
size will need to be tuned for each dataset once an appropriate
dimensionality has been chosen.

Figure \ref{results-parallel-batches} shows how the score is affected
by processing multiple batches in parallel. As long as the total
number of data points being trained simultaneously (the batch size
times the number trained in parallel) is much less than the total
dataset size, the score is not significantly affected.

\begin{table}[t]
  \begin{center}
  \begin{filecontents*}{systems.csv}
System,Precision at 10,Standard error
Random baseline,0.0056,0.0008
Global top items,0.0276,0.0021
Content,0.0281,0.0021
Collaborative,0.0750,0.0043
\end{filecontents*}
\csvautotabular{systems.csv}
  \end{center}
  \caption{Precision at 10 on the held out test set for the systems
    and baselines we considered.}
  \label{results-systems}
\end{table}

\subsection{Performance}

Tests were created to verify the different performance levels of the
Recommendation Engine in terms of the number of concurrent clients,
how old the recommended assets could be and how many containers were
deployed at the same time (horizontal scalability). We selected Jmeter
as the tool to run these tests given its simplicity for creating
different scenarios.

\begin{table}
  \begin{center}
  \begin{filecontents*}{performance.csv}
System,Users/sec,Average latency (ms)
Collaborative,663,382
Content,412,690
Top articles,305,86
\end{filecontents*}
\csvautotabular{performance.csv}
  \end{center}
  \caption{Real-time performance results for our system using eight
    containers for each recommendation algorithm and thirty concurrent
    clients.}
  \label{results-systems}
\end{table}

Results are shown in table \ref{results-systems}. Latency for
content-based recommendations is higher because the model size (sparse
vectors with up to 5000 non-zero dimensions) is much greater than for
the collaborative system (in this test the vectors were 200
dimensional dense vectors); the majority of the time is spent
transferring the model from BigTable to the cluster. Top article
latency is low because the top articles can be cached between
subsequent requests, and all that is needed is to retrieve and exclude
seen items for the user.


\section{Architecture}
\label{section:architecture}

Our focus on the architecture was to provide a base structure where
the collaborative and content based algorithms had enough resources to
ingest and train the data with minimum delay and maximum
scalability. Although intended to be used initially on \emph{The
  Times} and \emph{The Sun}, the engine needed to be product agnostic
and with delivery model of Software as a Service (SaaS), where a
single deployed version of the application is used for all customers.

Given these requirements, each component had to be designed to be
horizontally scalable to support huge amounts of data without
increasing the recommendation serving latency and freshness. For that,
we used a containerised architecture with Docker and Kubernetes to
allow easy control of the scalability. Data stores would also have to
cope with these variations on the amount of data and be able to scale
up and down.

PubSub was used for data ingestion since we can guarantee that all
messages are processed using its acknowledgment model and we can
configure its bandwidth quota as needed, although the default of 100
mb/s should be sufficient.  Bigtable was selected as the main storage
for being a massively scalable NoSQL database with low latency and
high throughput. It stores all the user actions, user models and asset
models.

Clients of the engine typically will query for the latest
recommendations, including the last few hours or days, but we also
wanted to support cases where they would need recommendations for much
longer periods. Because of that, it was not feasible to load thousands
of models from Bigtable, which sits outside the cluster and has much
more limited bandwidth compared with the cluster's internal speed. We
created a cache component that would run inside the cluster and
initialise all existing recommendable assets models from
Bigtable. Each serving API also had its own short-lived in-memory
cache for further optimisation.

\begin{figure}[t]
  \begin{center}
    \includegraphics[scale=0.5]{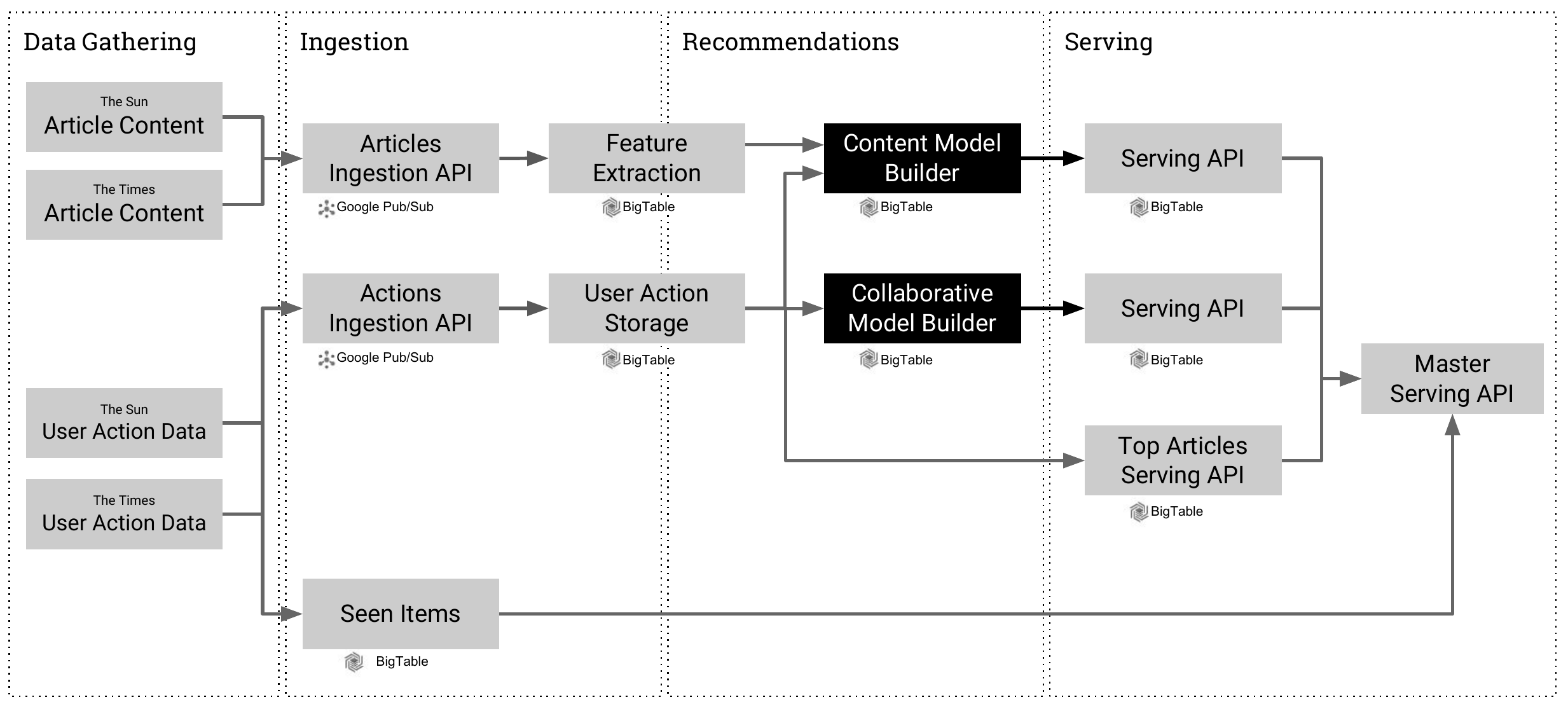}
  \end{center}
  \caption{The major components of our real-time architecture.}
  \label{figure:architecture}
\end{figure}

All the components of the Recommendation Engine can be divided into
four layers (see Figure \ref{figure:architecture}):
\begin{itemize}
  \item Data gathering: collection of user actions and content from
    our online publications to send to PubSub.
  \item Data ingestion: messages from PubSub are processed and stored
    in the engine. User actions are stored in BigTable and feature
    extraction is performed on article content.
  \item Data training: use collaborative and content based algorithms to
    train the ingested data and store the resulting models.
  \item Recommendation serving - APIs to generate recommendations using
    query parameters and previously generated assets models and user
    models.
\end{itemize}

New algorithms can be incorporated into the Recommendation Engine by
creating components fitting into the training and recommendation
serving layers.



\section{Conclusion}
\label{section:conclusion}

We have described the recommendation system currently used in
production at News UK. We make use of a novel algorithm for
incrementally updating collaborative filtering models. We demonstrated
its effectiveness in an offline evaluation and described the
conditions under which the incremental update works reliably for our
dataset.

In future work, we hope to combine our content-based and collaborative
systems. In our ongoing online tests measuring click-through rates on
recommendations, we have found that for some users, content-based
recommendations seem to be more effective, while for others, the
collaborative filtering recommendations give higher click-through
rates. We would like to be able to give the best recommendations
possible to each user, so we may try and learn which system works best
for users, perhaps using a multi-armed bandit approach. We will also
investigate hybrid recommendation techniques.
\
\section*{Acknowledgments}

Many thanks to Dan Gilbert and Jonathan Brooks-Bartlett for
feedback and support.

\bibliography{recs}{}

\begin{thebibliography}{10}
\providecommand{\url}[1]{\texttt{#1}}
\providecommand{\urlprefix}{URL }

\bibitem{tensorflow2015whitepaper}
Abadi, M., Agarwal, A., Barham, P., Brevdo, E., Chen, Z., Citro, C., Corrado,
  G.S., Davis, A., Dean, J., Devin, M., Ghemawat, S., Goodfellow, I., Harp, A.,
  Irving, G., Isard, M., Jia, Y., Jozefowicz, R., Kaiser, L., Kudlur, M.,
  Levenberg, J., Man\'{e}, D., Monga, R., Moore, S., Murray, D., Olah, C.,
  Schuster, M., Shlens, J., Steiner, B., Sutskever, I., Talwar, K., Tucker, P.,
  Vanhoucke, V., Vasudevan, V., Vi\'{e}gas, F., Vinyals, O., Warden, P.,
  Wattenberg, M., Wicke, M., Yu, Y., Zheng, X.: {TensorFlow}: Large-scale
  machine learning on heterogeneous systems (2015),
  \url{https://www.tensorflow.org/}, software available from tensorflow.org

\bibitem{covington2016deep}
Covington, P., Adams, J., Sargin, E.: Deep neural networks for youtube
  recommendations. In: Proceedings of the 10th ACM Conference on Recommender
  Systems. New York, NY, USA (2016)

\bibitem{diaz2012real}
Diaz-Aviles, E., Drumond, L., Schmidt-Thieme, L., Nejdl, W.: Real-time top-n
  recommendation in social streams. In: Proceedings of the sixth ACM conference
  on Recommender systems. pp. 59--66. ACM (2012)

\bibitem{fan2008liblinear}
Fan, R.E., Chang, K.W., Hsieh, C.J., Wang, X.R., Lin, C.J.: Liblinear: A
  library for large linear classification. Journal of machine learning research
   9(Aug),  1871--1874 (2008)

\bibitem{hu2008collaborative}
Hu, Y., Koren, Y., Volinsky, C.: Collaborative filtering for implicit feedback
  datasets. In: Data Mining, 2008. ICDM'08. Eighth IEEE International
  Conference on. pp. 263--272. Ieee (2008)

\bibitem{liu2009learning}
Liu, T.Y., et~al.: Learning to rank for information retrieval. Foundations and
  Trends{\textregistered} in Information Retrieval  3(3),  225--331 (2009)

\bibitem{pilaszy2010fast}
Pil{\'a}szy, I., Zibriczky, D., Tikk, D.: Fast als-based matrix factorization
  for explicit and implicit feedback datasets. In: Proceedings of the fourth
  ACM conference on Recommender systems. pp. 71--78. ACM (2010)

\bibitem{ponsford2017}
Ponsford, D.: {NRS} national press readership data: {Telegraph} overtakes
  {Guardian} as most-read `quality' title in print/online. Press Gazette  (26
  June, 2017)

\bibitem{siddiqui2014xstreams}
Siddiqui, Z.F., Tiakas, E., Symeonidis, P., Spiliopoulou, M., Manolopoulos, Y.:
  xstreams: Recommending items to users with time-evolving preferences. In:
  Proceedings of the 4th International Conference on Web Intelligence, Mining
  and Semantics (WIMS14). p.~22. ACM (2014)

\bibitem{takacs2011applications}
Tak{\'a}cs, G., Pil{\'a}szy, I., Tikk, D.: Applications of the conjugate
  gradient method for implicit feedback collaborative filtering. In:
  Proceedings of the fifth ACM conference on Recommender systems. pp. 297--300.
  ACM (2011)

\bibitem{zhou2008large}
Zhou, Y., Wilkinson, D., Schreiber, R., Pan, R.: Large-scale parallel
  collaborative filtering for the netflix prize. In: International Conference
  on Algorithmic Applications in Management. pp. 337--348. Springer (2008)

\end{thebibliography}
\bibliographystyle{splncs03}

\end{document}